\documentstyle[12pt,fleqn]{article}

\title{Amplification of Gamma Radiation from X-Ray Excited Nuclear States
\footnote{Published in Revue Roumaine de Physique {\bf 27}, 559 (1982)}}

\author{S. Olariu,
\thanks{Present address: 
Silviu Olariu, Institute of Physics and Nuclear 
Engineering, Department of Fundamental Experimental Physics, 76900 Magurele, 
P.O. Box MG-6, Bucharest, Romania; e-mail: olariu@ifin.nipne.ro}
$\:$I. Iovitzu Popescu\\
Central Institute of Physics, Bucharest, Magurele, Romania\\
\and
C. B. Collins\\
Center for Quantum Electronics, University of Texas at Dallas, \\
Box 688, Richardson, Texas 75080, USA}

\begin{document}
\date{}
\maketitle
\abstract
In this paper we discuss the possibility of the excitation of nuclear 
electromagnetic transitions by the absorption of X-ray quanta produced in 
appropriate inner-shell atomic transitions, and the relevance of this process
for the amplification of the gamma radiation from the excited nuclear states.
It is concluded that the X-ray pumping technique might provide a useful
approach for the development of a gamma ray laser.
\endabstract

\newpage

\section{Introduction}

The observed trend $[1,2]$ toward short-wavelength coherent sources of 
electromagnetic radiation from stimulated atomic transitions is presently
confronted with difficulties inherent to the shorter atomic life-times of the
excited states, smaller cross sections for the stimulated emission, and the
lack of resonators appropriate at these frequencies. On the other hand, because
of their small dimensions, the atomic {\it nuclei} have comparatively
long-lived excited states, while the Breit-Wigner cross section for the
interaction with the electromagnetic radiation is essentially the same for
atoms and nuclei at a given wavelength. Considerable effort has been therefore
directed toward the development of stimulated emission devices which would use
the M\"{o}ssbauer effect in nuclear transitions for the generation of
electromagnetic radiation in the 1 \dots 100 keV range of photon energy.
The difficulties in the development of the coherent gamma ray sources and the
perceived reward of success have been recently reviewed by Baldwin, Solem and
Gol'danskii $[2]$ with the general conclusion that while no scientific
principle has yet been shown to prohibit gamma ray lasers, further progress in
this direction is dependent upon research and technological advances in many
areas. 

Most of the earlier proposals for a gamma ray laser $[3-12]$ involved 
neutrons, either for {\it in situ} pumping $[5,7-9]$ or for the production of
long-lived nuclear isomers $[10,6,11,12]$. However, the relatively low
intensities of the available fluxes of neutrons and the difficulties inherent
in the narrowing of the effective widths in the case of the isomers $[13-17]$
create major obstacles along these lines. In this context attention was given
to the excitation of nuclear electromagnetic transitions, and among the
processes considered are the use of bremsstrahlung radiation $[18]$,
characteristic X radiation $[19,20]$, resonant M\"{o}ssbauer radiation
$[9,21,22]$, optical laser radiation $[23-31]$, and synchrotron radiation
$[32,33]$. 

In this paper we shall discuss the possibility of the excitation of nuclear
electromagnetic transitions by the absorption of X-ray quanta produced in the
inner-shell atomic transitions, and the relevance of this process for the
amplification of the gamma radiation from the exited nuclear states. It is
shown that a significant level of nuclear excitation can be obtained by an
appropriate choice of the atomic X-ray transition. The X-ray power required for
the pumping of a gamma ray laser is compared with the parameters of existing
X-ray flash devices. The nuclides whose level structure appears to be favorable
for the gamma-ray amplification are tabulated together with the X-ray pumping
transitions. It is concluded that the X-ray flash pumping technique might
provide a useful approach for the development of a gamma ray laser, and
motivated further investigation of the process of excitation of nuclear
electromagnetic transitions.

\section{Basic concepts}

According to the general quantum mechanical description, the atomic nucleus can
exist in a series of stationary states characterized by their energy, spin and
parity. If the nucleus is in one of its excited states it generally undergoes a
transition to a lower state which is accompanied by a corresponding energy
transfer to the radiation field, atomic electrons, or other particles. When the
transition energy is converted into a gamma-ray photon, a small amount of the
total energy appears in the final state as kinetic energy of the whole nucleus.
Since this recoil energy is generally large compared to the nuclear level
width, the gamma-ray emission and absorption lines of the free nuclei are
shifted and broadened, and the cross section for the resonant interaction is 
correspondingly reduced. This difficulty can be avoided by making use of the
M\"{o}ssbauer effect $[34]$ for nuclei bound in a crystalline lattice for
transition energies which do not exceed 100 keV

The dominant multipolarity of low gamma-ray transitions is M1, although E1 and
E2 transitions also occur, and the typical life-time of these states is in the
nanosecond range. The conventional method for the excitation of low-lying
nuclear states is through beta decay or orbital electron capture from the
contiguous nuclides. However, since the life-times of the decaying nuclei are
very much longer than the life-times of the excited states, the concentration
of the latter is extremely small. Another excitation technique would be the
irradiation with a flux of neutrons, but again impractically large neutron
fluxes are required to obtain significant population of the nanosecond excited
states. 

On the other hand, the excitation of the nuclear states with electromagnetic
radiation of appropriate energy appears to produce a significant level of
population in the upper state. The electromagnetic radiation can be produced in
X-ray transitions between the states of the atomic electron shell, and since
the X-ray line spectrum involves energies up to about 100 keV for the heavy
elements it seems appropriate for the excitation of the M\"{o}ssbauer
transitions. The width of the X-ray states are 0.1\dots 1 eV, and therefore the
very short-lived atomic states are not suitable for direct amplification of the
radiation.

\section{Amplification of gamma radiation}

Provided that a reasonable matching between the nuclear and the atomic
transition energies could be obtained, an X-ray flash might be used to raise
the nuclei from the ground state, $a$, to an upper state, $b$, in a process
analogous to the optical pumping of the atomic transitions (Fig. 1). 
Now the process of
transition to a lower state, $c$, through the emission of a gamma ray photon
would be greatly enhanced in the presence of a large population in the upper
state, $b$. As pointed out by several authors $[2]$ the development of a gamma
ray pulse would require the population inversion between the states $b$ and
$c$, and also the resonant gain must exceed the non-resonant losses.

Since it is generally accepted that the gamma ray laser would be a single pass
device $[2]$, the number of photons in the cascade induced by the spontaneously
emitted gamma ray quanta must be of the order of the total number of nuclei in
the upper state $b$,
\begin{equation}
{e^{\sigma_{bc}n_b L}}\leq  n_b L a^2 ,
\end{equation}
where $\sigma_{bc}$ is the cross section for the transition 
$b\rightarrow c, n_b$ is the concentration of nuclei in the state $b$, $L$ is 
the length and $a$ is the transverse dimension of the nuclear sample. Moreover,
in order to have negligible diffraction losses, it is necessary $[2]$ that
\begin{equation}
L\lambda \leq \frac{1}{3} a^2 ,
\end{equation}
where $\lambda $ is the gamma ray wavelength.

The conditions, Eqs. (1) and (2), define the upper state concentration $n_b$
and the area, $a^2$, as functions of the length $L$ of the nuclear sample. 
Increasing the length $L$ results in a larger transverse dimension $a$, but 
generally in a lower threshold {\it density} of the nuclei in the upper state.
However, the total number $N_b=n_b L a^2$ of nuclei in the upper state
increases with $L$, provided that $N_b\sigma_{bc}>L\lambda$. On the other hand,
increasing the wavelength $\lambda$ for fixed dimensions $L, a$ results in
lower threshold values for the concentration $n_b$ and the total number of
excited nuclei $N_b$.

\section{The X-ray pumping of nuclear electromagnetic transitions}

The building up of a large concentration of nuclei in the excited state $b$ is
facilitated by the fact that the atomic electrons screen the nucleus from
interactions which otherwise would lead to the broadening of lines. In fact,
although the concentration of nuclei in the upper state turns out to be large
compared with typical inversion densities at optical frequencies, it represents
a small fraction of the ground state concentration. The probability, $w$, for
the excitation of a nucleus by an X-ray pulse on ${\cal N}_x$ quanta per cm$^2$
having a spectral width $\Gamma_x$ around the nuclear transition energy is
\begin{equation}
w=\sigma_{ab}{\cal N}_x \frac{\Gamma_b}{\Gamma_x} ,
\end{equation} 
where $\sigma_{ab}$ represents the cross section for resonant nuclear
excitation and $\Gamma_b$ is the width of the upper state $b$. In general, the
X-ray lines are broad compared to the widths of the nuclear levels, and only a
small fraction $\Gamma_b/\Gamma_x$ of the X-ray quanta is effective in the
excitation process. On the other hand, the relatively large X-ray width
facilitates the matching of the resonance condition between the X and the gamma
transitions.

Since it is the {\it density} of the X-ray photons which determines the pumping
rate, Eq. (3), it seems that a suitable X-ray source would be a filamentary
plasma in the immediate proximity of the nuclear sample, as represented in Fig.
2. Among the devices which produce such flashes are the vacuum sparks, the
exploding wires, the plasma focus, and the laser focus $[36]$.

\section{Case study and tabulation}

The minimum fluence ${\cal E}$ of the X-ray energy that must be injected into
the sample is proportional to the nuclear transition energy $E_{ab}$, to the
population of the upper state, $N_b$, and to the ratio $\Gamma_x/\Gamma_b$ of
the atomic and nuclear widths:
\begin{equation}
{\cal E}=E_{ab} N_b\frac{\Gamma_x}{\Gamma_b} .
\end{equation}
The quantity ${\cal E}$ is represented in Fig. 3 for values of the parameters
likely to be encountered for gamma ray devices. The total energy in the X-ray
flash must probably be at least one order of magnitude higher than that
represented in Fig. 3 because of solid angle problems. For comparison, the
presently available X-ray devices provide flashes with a duration as short as
10 ps and X-ray powers in excess of $10^{10}$ watts $[37]$, thus being able to
provide about $10^{-1}$ Joules in periods of time which are short relative to
the nuclear life-times.

Assuming that the cross section for stimulated emission of the gamma rays is
$10^{-18}$ cm$^2$, the length $L$ of the nuclear sample of 1 cm, then in order
to obtain the significant gain of 1/cm it is necessary a concentration of
$n_b=10^{18}$ nuclei/cm$^3$. Since according to Eq. (2), $a=10^{-4}$ cm, the
number of excited nuclei is $N_b=3\times 10^{10}$. Therefore, the energy stored
in the nuclear excitation is about $6\times 10^{-6}$ Joules, and the energy of
the gamma ray pulse would be of the order of $10^{-6}$ Joules. The intensity of
the gamma ray pulse at 10 cm from the source would be of the order of $10^{9}$
watts/cm$^2$, or about $10^{14}$ Ci/cm$^2$.

Further tabulated are those nuclides which appear to be of interest for the
electromagnetic excitation with atomic X rays.

\section{Conclusions}

We have shown in this paper that the X-ray excitation of nuclear
electromagnetic transitions might provide a technique for the pumping of a
gamma ray laser. The performances of existing pulsed X-ray sources are one or
two orders of magnitude below the threshold for the gamma amplification.
Additional investigation is needed to ensure that the nuclear sample is stable
enough against the high intensity X-ray pulse. This is consistent with the
general conclusion that further progress in this direction is dependent upon
research and technological advances in many areas.

\vspace*{1cm}

{\footnotesize {\it Acknowledgment}. 
This paper was supported in part by the National Science Foundation 
under Grant No. INT 76-18982 and in part by the Romanian State Committee for 
Nuclear Energy under the U.S.-Romanian Cooperative Program in Atomic and 
Plasma Physics.}
\newpage

\begin{center}
REFERENCES
\end{center}

\noindent
1. D. J. Nagel, Phys. Fenn. {\bf 9}, 381 (Supplement 1)$\:$(1974).\\
2. G. C. Baldwin, J. C. Solem and V. I. Gol'danskii, Rev. Mod. Phys. 
{\bf 53}, 687 (1981).\\
3. G. C. Baldwin, J. P. Neissel, J. H. Terhune, and L. Tonks, Trans. Am. Nucl.
Soc. {\bf 6}, 178 (1963).\\
4. L. A. Rivlin, Vopr. Radioelekytron. {\bf 6}, 42 (1963).\\
5. V. I. Gol'danskii and Yu. Kagan, Zh. Eksp. Teor. Fiz., {\bf 64}, 90
(1973)$\:$(Sov. Phys.-JETP, {\bf 37}, 49 (1973)).\\
6. V. S. Letokhov, Zh. Eksp. Teor. Fiz. {\bf 64}, 1555 (1973)$\:$
(Sov. Phys.-JETP, {\bf 37}, 787 (1973)).\\
7. J. C. Solem, Los Alamos Scientific Laboratory, Report No. LA-7898-MS 
(1979).\\
8. L. Wood and G. Chapline, Nature {\bf 252}, 447 (1974).\\
9. V. I. Gol'danskii, Yu. Kagan and V. A. Namiot, Zh. Eksp. Teor. Fiz. Pis'ma
Red., {\bf 18}, 34 (1973)$\:$(JETP Lett. {\bf 18}, 34 (1973)).\\
10. G. C. Baldwin, J. P. Neissel and L. Tonks, Proc. IEEE, {\bf 51}, 1247
(1963). \\
11. V. I. Gol'danskii and Yu. M. Kagan, Usp. Fiz. Nauk, {\bf 110}, 445 (1973)
(Sov. Phys.-Usp. {\bf 16}, 563 (1974)).\\
12. V. Vali and M. Vali, Proc. IEEE, {\bf 51}, 182 (1963).\\
13. R. V. Khokhlov, Zh. Eksp. Teor. Fiz. Pis'ma Red. {\bf 15}, 580 (1972)$\:$
(JETP Lett., {\bf 15}, 414 (1972).\\
14. Yu. A. Il'inskii and R. V. Khokhlov, Usp. Fiz. Nauk, {\bf 110}, 448
(1974)$\:$(Sov. Phys.-Usp. {\bf 16}, 565 (1974)). \\
15. V. I. Gol'danskii, S. V. Karyagin and V. A. Namiot, Zh. Eksp. Teor. Fiz. 
Pis'ma Red., {\bf 19}, 625 (1974) (JETP Lett. {\bf 19}, 324 (1974)).\\
16. V. A. Namiot, Zh. Eksp. Teor. Fiz. Pis'ma Red. {\bf 18}, 369 (1973)$\:$
(JETP Lett., {\bf 18}, 216 (1973)).\\
17. S. V. Karyagin, Zh. Tekh. Fiz. Pis'ma {\bf 2}, 500 (1976)$\:$
(Sov. Phys.-Tech. Phys. Lett., {\bf 2}, 196 (1976)).\\
18. D. Marcuse, Proc IEEE, {\bf 51}, 849 (1963).\\
19. V. S> Letokhov, Kvantovaya Elektron., {\bf 4}, 125 (1973)$\:$
(Sov. J. Quantum Electr., {\bf 3}, 360 (1974)).\\
20. V. I. Vysotskii, Zh. Eksp. Teor. Fiz. {\bf 77}, 492 (1979)$\:$
(Sov. Phys.-JETP, {\bf 50}, 250 (1979)).\\
21. G. C. Baldwin and J. C. Solem, J. Appl. Phys. {\bf 51}, 2372 (1980).\\
22. S. V. Karyagin, Zh. Eksp. Teor. Fiz., {\bf 79}, 730 (1980)
$\:$(Sov. Phys.-JETP, {\bf 52}, 370 (1980)).\\
23. J. W. Eerkens, U.S. Patent 3,430,046 (1969).\\
24. E. V. Bakhlanov and P. V. Chebotaev, Zh. Eksp. Teor. Fiz. Pis'ma Red 
{\bf 21}, 286 (1975)$\:$(JETP Lett. {\bf 21}, 131 (1975)).\\
25. L. A. Rivlin, Kvantovaya Elektron., {\bf 4}, 676 (1977)$\:$(Sov. J. Quantum
Electron., {\bf 7}, 380 (1977)).\\
26. C. B. Collins, S. Olariu, M. Petrascu and I. Iovitzu Popescu, Phys. Rev.
Lett., {\bf 42}, 1379 (1979).\\
27. C. B. Collins, S. Olariu, M. Petrascu and I. Iovitzu Popescu, Phys. Rev. C,
{\bf 20}, 1942 (1979).\\
28. B. Arad, S. Eliezer and A. Paiss, Phys. Lett. A, {\bf 74}, 395 (1979).\\
29. S. Olariu, I. Iovitzu Popescu and C. B. Collins, Phys. Rev. C, {\bf 23}, 
50 (1981).\\
30. S. Olariu, I. Iovitzu Popescu and C. B. Collins, Phys. Rev. C, {\bf 23},
1007 (1981).\\
31. C. B. Collins, F. W. Lee, D. M. Shemwell, B. D. DePaola, S. Olariu, and I.
Iovitzu Popescu, J. Appl. Phys., (1982).\\
32. V. F. Dmitriev and E. V. Shuryak, Zh. Eksp. Teor. Fiz., {\bf 67}, 494
(1974)$\:$(Sov. Phys. -JETP, {\bf 40}, 244 (1975)).\\
33. R. L. Cohen, G. L. Miller, K. H. West, Phys. Rev. Lett., {\bf 41}, 381
(1978).\\ 
34. R. L. M\"{o}ssbauer, Z. Physik, {\bf 151}, 124 (1958).\\
35. L. Allen and G. I. Peters, J. Appl. Phys., A, {\bf 4}, 564 (1971).\\
36. D. J. Nagel, in {\it Advances in X-ray Analysis}, vol. 18, edited by W. L.
Pickles, C. S. Barrett, J. B. Newkirk and C. O. Rund, Plenum, 1974, p. 1.\\
37. D. J. Nagel and C. M. Dozier, Proc. 12$^{\rm th}$ Int. Congr. High Speed
Photography, Toronto, Canada, 1976, in Soc. Photo-Optical Instrum. Engrs.,
1977, p. 2.\\

\newpage
\begin{table}
\caption{Nuclear transitions which can be excited with atomic X-rays (Compiled
from C. M. Lederer and V. S. Shirley, {\it Table of Isotopes}, 7$^{\rm th}$
edition, J. Wiley \& Sons, 1978)}
\vspace*{1cm}
\begin{center}
\begin{tabular}{|c|c|c|c|c|c|}
\hline
\multicolumn{3}{|c|}{Nuclear transition} & \multicolumn{2}{c|}{Atomic X-ray} &
Stimulated \\
\multicolumn{3}{|c|}{to the upper state} & \multicolumn{2}{c|}{transition} &
transition \\
\hline
Element & Transition & Ground state & Element & Transition and & Energy, keV\\
        & energy, keV & life-time   &         & energy, keV    &            \\
\hline
$^{66}$Ga & 66.32 & 9.4 h & $_{79}$Au & 66.369, $KL_I$ & 22.43\\
$^{100}$Rh & 74.8 & 20.8 h & $_{83}$Bi & 74.815, $KL_{II}$ & 42.1\\
$^{140}$La & 43.81 & 40.3 h & $_{61}$Pm & 43.821, $KM_{II}$ & 13.85\\
$^{143}$Ce & 42.3 & 33.0 h & $_{64}$Gd & 42.308, $KL_{II}$ & 23.4\\
$^{144}$Pm & 66.63 & 349.1 d & $_{74}$W & 66.706, $KM_{I}$ & 5.89\\
$^{151}$Sm & 69.69 & 87.9 y & $_{75}$Re & 69.726, $KM_{IV}$ & 64.87\\
$^{161}$Dy & 43.84 & stable & $_{61}$Pm & 43.821, $KM_{IV}$ & 18.19\\
$^{162}$Tm & 66.9 & 21.7 m & $_{79}$Au & 66.989, $KL_{II}$ & 22.2\\
$^{165}$Er & 62.68 & 10.3 h & $_{77}$Ir & 62.692, $KL_{I}$ & 15.52\\
$^{167}$Yb & 33.91 & 17.5 m & $_{58}$Ce & 33.895, $KL_{I}$ & 4.25\\
$^{175}$Ta & 51.38 & 10.5 h & $_{70}$Yb & 51.354, $KL_{II}$ & 14.98\\
$^{191}$Pt & 30.36 & 2.9 d & $_{55}$Cs & 30.272, $KL_{I}$ & 20.80\\
$^{201}$Hg & 32.19 & stable & $_{56}$Ba & 32.194, $KL_{II}$ & 30.61\\
$^{223}$Ra & 61.52 & 11.4 d & $_{76}$Os & 61.486, $KL_{II}$ & 31.61\\
$^{237}$U & 56.3 & 6.7 d & $_{73}$Ta & 56.277, $KL_{II}$ & 44.9\\
$^{245}$Am & 28.0 & 2.0 h & $_{53}$I & 27.982, $KL_{I}$ & 8.8\\
$^{246}$Am & 43.81 & 25.0 m & $_{65}$Tb & 43.742, $KL_{II}$ & 27.6\\
\hline
\end{tabular}
\end{center}
\end{table}

\end{document}